# Anisotropic Dielectric Breakdown Strength of Single Crystal Hexagonal Boron Nitride

Yoshiaki Hattori*[†], Takashi Taniguchi[‡], Kenji Watanabe[‡] and Kosuke Nagashio**[†§]
[†]Department of Materials Engineering, The University of Tokyo, Tokyo 113-8656, Japan
[‡]National Institute of Materials Science, Ibaraki 305-0044, Japan
[§]PRESTO, Japan Science and Technology Agency (JST), Tokyo 113-8656, Japan
*hattori@adam.t.u-tokyo.ac.jp, **nagashio@material.t.u-tokyo.ac.jp

**ABSTRACT:** Dielectric breakdown has historically been of great interest from the perspectives of fundamental physics and electrical reliability. However, to date, the anisotropy in the dielectric breakdown has not been discussed. Here, we report an anisotropic dielectric breakdown strength ($E_{BD}$) for $h$-BN, which is used as an ideal substrate for two-dimensional (2D) material devices. Under a well-controlled relative humidity, $E_{BD}$ values in the directions both normal and parallel to the $c$ axis ($E_{BD\perp c}$ & $E_{BD//c}$) were measured to be 3 and 12 MV/cm, respectively. When the crystal structure is changed from sp$^3$ of cubic-BN ($c$-BN) to sp$^2$ of $h$-BN, $E_{BD\perp c}$ for $h$-BN becomes smaller than that for $c$-BN, while $E_{BD//c}$ for $h$-BN drastically increases. Therefore, $h$-BN can possess a relatively high $E_{BD}$ concentrated only in the direction parallel to the $c$ axis by conceding a weak bonding direction in the highly anisotropic crystal structure. This explains why the $E_{BD//c}$ for $h$-BN is higher than that for diamond. Moreover, the presented $E_{BD}$ value obtained from the high quality bulk $h$-BN crystal can be regarded as the standard for qualifying the crystallinity of $h$-BN layers grown via chemical vapor deposition for future electronic applications.

## INTRODUCTION

Dielectric breakdown in solids is a long-known and well-investigated phenomenon. The electron avalanche mechanism was first proposed by A. von Hippel[1] and was quantum-mechanically formulated by H. Fröhlich.[2] Interestingly, the dielectric breakdown of mica, which is a layered material with a thickness of as little as ~20 nm, [3,4] was reported to be ~10 MV/cm as early as in the 1950s and 60s. However, the breakdown of mica in the direction parallel to the basal plane was not investigated, primarily because of the difficulty of fabricating two separate electrodes with a sufficiently narrow gap for the breakdown. Since then, because of the technological importance of the dielectric breakdown of thin $SiO_2$ films, intensive and detailed investigations of phenomena such as the effects of impurities and defects captured in $SiO_2$ during the oxidation of Si and the application of electrical stress to a $SiO_2$/Si interface have been performed. [5,6] However, the anisotropy of dielectric breakdown has not been extensively discussed because the gate oxide should be amorphous to minimize the leakage path. To date, there have been only a few reports on such anisotropy, for example, the existence of a preferred breakdown direction in alkali-halide crystals[1,7] and the anisotropic $E_{BD}$ observed in SiC power device applications.[8] Thus, currently, the experimental data are limited.

Here, we study single-crystal hexagonal boron nitride ($h$-BN), which is a wide-band-gap insulator (5.2 – 5.9 eV) with a highly anisotropic layered structure. It is widely utilized as a substrate and gate insulator to achieve high carrier mobility in layered channel materials such as graphene,[9] MoS$_2$,[10] black phosphorus,[11] and others because its atomically flat surface without dangling bonds results in considerably reduced scattering at the graphene/substrate interface. More importantly, for the $h$-BN/graphene/$h$-BN heterostructure, a mobility of higher than ~100,000 cm$^2$/Vs can surprisingly be achieved "at room temperature".[12] The main limiting factor for the mobility in graphene is remote phonon scattering from the $h$-BN substrate, which is weak due to the high surface optical phonon energy of 101.7 meV compared with that for conventional $SiO_2$ substrate (58.9 meV).[13-16] The origin of this high phonon energy relies on the strong B-N bonding, suggesting a higher breakdown strength. Therefore, from the perspectives of device application and fundamental physics, a quantitative investigation of the dielectric breakdown of $h$-BN is important because the insulating properties of gate insulators such as leakage current and electrical reliability are directly related to the transistor performance. To date, several researchers have measured the dielectric breakdown strength of $h$-BN bulk single crystals in the direction parallel to the $c$ axis ($E_{BD//c}$), obtaining values of ~10 – 12 MV/cm.[17-19] Interestingly, this value is either comparable to or is slightly higher than the ideal value of ~10 MV/cm for diamond.[20-22] Moreover, we have observed layer-by-layer breakdown behavior,[19] suggesting the existence of a strong anisotropy in the dielectric breakdown originating from the layered structure. Anisotropy in various properties, such as thermal conductivity,[23] thermal diffusion,[24] thermal expansion,[25] magnetic susceptibility,[26] and polarization,[27] has been reported for $h$-BN. Anisotropy in $E_{BD}$, however, has not been revealed yet, even though $h$-BN can be regarded as the ideal model material for such anisotropy.

In this study, we present a complete characterization of the dielectric breakdown strength of high-quality $h$-BN bulk single crystals in the directions both normal and parallel to the $c$ axis ($E_{BD\perp c}$ & $E_{BD//c}$) to reveal the overall picture of the anisotropy in the dielectric breakdown of this material.

**EXPERIMENTAL SECTION**
**Device fabrication.** Figures 1a,b show a schematic drawing and an optical image of a typical $h$-BN device fabricated by means of electron beam (EB) lithography[28] and used for $E_{BD\perp c}$ measurements. The $E_{BD}$ values of wide-gap semiconductors such as GaN and SiC have been evaluated using a similar electrode structure for power device applications.[29,30] The design of this experimental setup addresses four important points. First, mechanical exfoliation of bulk $h$-BN single crystals grown at a high pressure and a high temperature was used in this study because the impurity density of C and O in such crystals is small,[31] and the crystal quality has been confirmed by many researchers. Second, high-quality synthetic fused silica (of ES grade for optical use) with a root-mean-square (RMS) roughness of 0.4 nm was used as the substrate because the $E_{BD}$ of the substrate must be higher than that of the $h$-BN. It should be noted that an $SiO_2$/Si substrate cannot be used for this purpose. Third, the $L_{gap}$ was chosen to be in the range of 200 nm to 1 μm to apply a parallel electrical field without air discharge. According to Paschen's law,[32] the maximum voltage of the source meter for $L_{gap}$ = 1 μm guarantees that no air discharge occurs. **Figure 1c** shows the calculated electrical potential in the form of a color map for $L_{gap}$ = 500 nm and for $V$ = 100 V at the cross-sectional plane indicated by the white dotted rectangle in **Fig. 1a** (details provided in **Supporting Fig. S1**). Although the electrical field within approximately 50 nm of the edge of the electrode is locally enhanced, as shown by the white dotted circle, the electrical field applied to the main $h$-BN body is entirely parallel to the in-plane direction. Therefore, the minimum $L_{gap}$ was chosen to be ~200 nm. Fourth, the relative humidity was carefully controlled using a humidifier and monitored by a hygrometer placed near the prober (**Supporting Fig. S2**). Although *h*-BN is hydrophobic,[33] $E_{BD}$ is expected to be strongly affected by the water molecules physisorbed onto the *h*-BN surface.

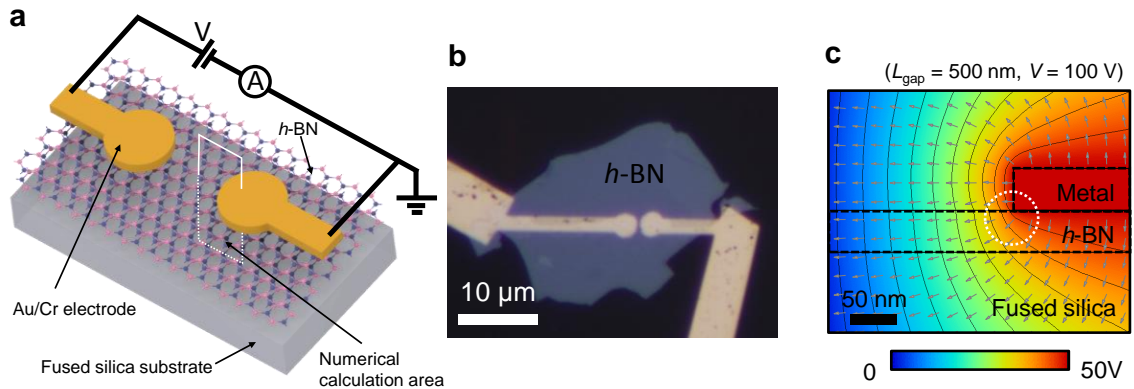

**Figure 1.** Device fabrication. (**a**) Schematic diagram of the measurement setup. (**b**) Optical microscopy image of a fabricated sample. (**c**) Intensity profile of the calculated electrical potential around the electrode. The solid lines and gray arrows represent the equipotential lines and the unit vectors of the electrical field, respectively. The tip of each electrode was rounded to avoid the microscopic electric field concentration observed at the corners of rectangular electrodes in general. However, the parallel plate electrodes can be assumed in the present device, because $L_{gap}$ is much shorter than the radius of the circular shape.

**Electrical measurements and characterization.** The electrode gaps were measured by atomic-force microscopy (AFM) in the tapping mode or scanning electron microscopy (SEM) in the low-vacuum mode. The thicknesses of the *h*-BN flakes were measured by AFM (typical measurement provided in **Supporting Fig. S3**). *I-V* measurements were performed in ambient air or vacuum (~5.0×10⁻³ Pa) at room temperature (21 – 25 °C) in a prober using a source meter (SMU 2450, Keithley). The maximum voltage supplied was 210 V. The voltage step and ramping rate were 0.050 V and 1.25 V/s, respectively. The substrate was electrically floating during the measurement. The compliance limit for the current was typically set to 1 µA. Cathodoluminescence (CL) spectra were measured using a spectrograph system (PDP-320, Photon Design) by SEM with a 5-kV accelerating voltage.

## RESULTS

**Dielectric Breakdown in the Direction Normal to the *c* Axis.** Before the $E_{BD\perp c}$ of the *h*-BN was investigated, the $E_{BD}$ of the fused silica substrate was evaluated with the same electrode structure but without *h*-BN at various relative humidities under atmospheric pressure and in a vacuum. The results of the typical *I-V* measurements are provided in **Supporting Fig. S4**. The dependence of the dielectric breakdown voltage ($V_{BD}$) on $L_{gap}$ is shown in **Fig. 2a**.

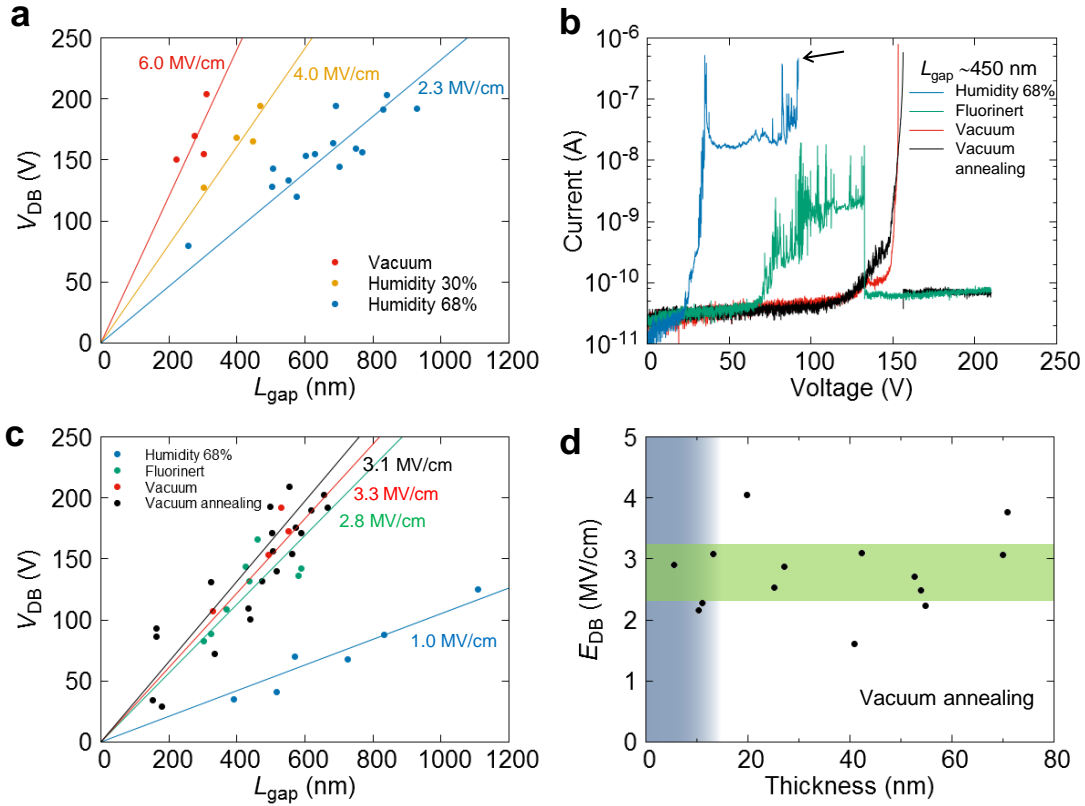

**Figure 2.** Dielectric breakdown in the direction normal to the *c* axis. (**a**) $V_{BD}$ as a function of $L_{gap}$ for the fused silica substrate. (**b**) *I-V* curves for *h*-BN with $L_{gap}$ ~450 nm on a fused silica substrate. (**c**) $V_{BD}$ as a function of $L_{gap}$ for *h*-BN on a fused silica substrate. Dielectric breakdown tests were conducted under various conditions to investigate the effect of the adsorbed water. Each color has a common interpretation in all four figures, **a-d**. The orange and blue lines represent tests performed under atmospheric pressure at relative humidities of 30 % and 68 %, respectively. The red lines represent tests performed under vacuum at ~5.0×10⁻³ Pa. The black lines represent vacuum annealing tests, in which the samples were annealed at 200 °C for two hours and then cooled to room temperature under vacuum before the breakdown test. The green lines represent breakdown tests conducted in Fluorinert, an insulating oil. (**d**) $E_{BD}$ as a function of the *h*-BN thickness.

The $E_{BD}$ value calculated from the slope of the $V_{BD}$ vs. $L_{gap}$ curve increases with decreasing relative humidity. The thicknesses of the adsorbed water layer on SiO$_2$ has been reported to be dependent on the relative humidity using attenuated total reflection infrared spectroscopy.[34] The differences in $E_{BD}$ observed in **Fig. 2a** can be attributed to the amount of water adsorbed on the fused silica substrate, which promotes the creeping discharge. When the adsorbed water was removed under vacuum, the $E_{BD}$ value further increased to ~6.0 MV/cm; this could be the intrinsic value for the present fused silica substrate. These results suggest that a sample with an $E_{BD}$ of less than 6.0 MV/cm can be evaluated on this fused silica substrate.

Once the effect of the adsorbed water on the dielectric breakdown had been confirmed, a more careful investigation of the $E_{BD\perp c}$ of $h$-BN itself was performed by comparing four types of experimental setups. **Figure 2b** shows the *I-V* measurements recorded for $L_{gap}$ ~450 nm and the $h$-BN thickness of 10 – 60 nm. When the relative humidity was ~68%, the current flow was quite unstable, exhibiting several peaks. $V_{BD}$ could be clearly defined as 92 V, as indicated by the arrow, because the electrode was broken by the melting and evaporation due to the high current density with perceivable light emission (**Fig. 3** and **Supporting Fig. S5**). The current fluctuation due to the adsorbed water is more evident than that observed for the fused silica substrate (**Supporting Fig. S4**). The water molecules physisorbed on $h$-BN surface may become unstable under a strong electric field, whereas the water molecules form hydrogen bonds with the OH terminations of a SiO$_2$ surface, resulting in a more stable leakage current behavior. Next, *I-V* measurements were performed under atmospheric pressure in Flourinert, an insulating oil (**Supporting Fig. S6**),[35,36] in accordance with the previous dielectric breakdown experiments on SiC and GaN.[37,38] The $V_{BD}$ value increased, and the leakage current before breakdown was reduced but was still present. When the adsorbed water was removed under vacuum, the leakage current decreased below the detectable level. Finally, the sample was annealed at 200 °C for two hours using the stage heater in the vacuum probe. The $V_{BD}$ value did not change from that observed in vacuum, suggesting that the effect of the adsorbed water was negligible. **Figure 2c** summarizes $V_{BD}$ as a function of $L_{gap}$ for $h$-BN on a fused silica substrate. The $E_{BD}$ value observed at the relative humidity of 68% was 1 MV/cm, whereas the other three cases yielded values of approximately 3 MV/cm; all of these values are smaller than that of the fused silica substrate. Based on these data, it can be concluded that for the investigated device structure, the value of $E_{BD\perp c}$ for $h$-BN is ~3 MV/cm. The effect of the electrical field concentration on $E_{BD\perp c}$ caused by the electrode shape is discussed in the **Supporting Note 1**.

Now, let us discuss the increase in the current just before breakdown. The results of repeated *I-V* measurements (see **Supporting Fig. S7** for details) show that the leakage current just before breakdown increased with the increasing number of measurements. This suggests the irreversible introduction of defects into the $h$-BN just before breakdown.

**Characterization of the Remaining $h$-BN Layers after Breakdown.** We analyzed the morphology of the $h$-BN after breakdown. **Figure 3a** shows an optical image of a typical sample after breakdown. Small particles with a diameter of less than 200 nm are evident in the electrode gap in the magnified AFM image of **Fig. 3b**. The results of energy-dispersive X-ray spectrometry (EDS) of these particles (**Supporting Fig. S8**) suggest that they were formed by the melting of the metal electrodes due to the high current density. Interestingly, as shown in **Fig. 3c**, the heights of the $h$-BN in the gap and below the electrode were reduced by ~4 nm, and the remaining $h$-BN surface was still flat. In the case of the breakdown of the fused silica substrate without the $h$-BN, the surface in the gap region became rough (**Supporting Fig. S9**). The flatness of the $h$-BN after breakdown appears to be a consequence of its layered structure. After many breakdown tests, the upper $h$-BN layers with the thickness of 4 – 15 nm were found to have melted.

Now let us consider the situation in which the initial $h$-BN thickness is smaller than that of the upper $h$-BN layers that will disappear during breakdown, and therefore, the contribution of the $h$-BN/substrate interface to the breakdown behavior should be taken into account. The relationship between $E_{BD\perp c}$ and the $h$-BN thickness is illustrated in **Fig. 2d**. For an $h$-BN thickness of less than 15 nm, the $h$-BN melted completely and the surface of the fused silica substrate was exposed after breakdown (**Supporting Fig. S10**). Despite the existence of an additional $h$-BN/substrate interface during the breakdown, the constant nature of the $E_{BD\perp c}$ data over the entire thickness range suggests the absence of any contribution from the $h$-BN/substrate interface. This result confirms the validity of the $E_{BD\perp c}$ value obtained for $h$-BN because all possible current paths (air or vacuum, $h$-BN surface, $h$-BN/substrate interface, substrate, and bottom floating stage) other than the *h*-BN itself are thoroughly considered.

The defect formation in the remaining *h*-BN layers after breakdown was analyzed via CL spectroscopy. **Figure 3d** shows the CL spectra at different positions, with their intensities normalized to that at 215 nm, which corresponds to the wavelength of free excitons in *h*-BN.[31] A broad peak near 225 nm appears in the fractured regions A and B, as indicated in **Fig. 3a**, unlike the regions C and D outside the gap. The CL mapping for the intensity ratio between 220 nm and 215 nm in **Fig. 3e** reveals that the broad peak near 225 nm clearly concentrates in the gap region. The bound exciton luminescence near 225 nm has been reported to be due to the stacking faults produced by a weak mechanical deformation.[39] Therefore, it is suggested that the *h*-BN in the gap region is deformed by the impact at dielectric breakdown. Although we expect some luminescence related to the defect levels formed in the energy gap of *h*-BN, these were not found.

Based on these results, the expected fracture process is schematically illustrated in **Fig. 4**. Avalanche breakdown is one of the mechanisms of dielectric breakdown.[40] With an increasing electrical field, the electrons injected into the *h*-BN through Fowler-Nordheim (FN) tunneling are accelerated by the electrical field and gain energy greater than the band gap to excite electron-hole (*e-h*) pairs, often resulting in bond breakage. The formation of the irreversible defects just before breakdown, which was confirmed by the irreversible *I-V* measurement **(Supporting Fig. 7)**, is indicated by open circles in **Fig. 4b**. Then, a large overcurrent flows through the sample at dielectric breakdown, at which time the metal electrodes and several upper layers of the *h*-BN are thermally melted because of the high current density. Simultaneously, the impact at dielectric breakdown creates stacking faults in the remaining *h*-BN layers, as shown in **Fig. 4c**.

**Dielectric Breakdown in the Direction Parallel to the *c* Axis.** Finally, the dielectric breakdown in the direction parallel to the *c* axis was investigated to evaluate the anisotropy of the dielectric breakdown of *h*-BN. **Figures 5a,b** show a schematic drawing and an optical image, respectively. *h*-BN with the thickness of 10 – 100 nm was sandwiched between the upper and lower Cr/Au electrodes using a transfer technique to form a metal-insulator-metal (MIM) structure (**Supporting Note 2**).[18,41,42] The overlap area was designed to be 25 μm². The conditions for the annealing and the *I-V* measurements were the same as those for the $E_{BD\perp c}$ tests. **Figure 5c** shows the typical *I-V* curve for an *h*-BN thickness of 14 nm in vacuum. The current increases rapidly (over less than 40 ms) at $V_{BD}$ = 22.7 V, whereas the effect of the adsorbed water is negligible. $E_{BD//c}$ is plotted as a function of the *h*-BN thickness in **Fig. 5d**. The $E_{BD//c}$ results exhibit a similar trend to those observed in our previous study using conductive AFM,[19] supporting the validity of the present measurement. The value of $E_{BD//c}$ was determined to be ~12 MV/cm.

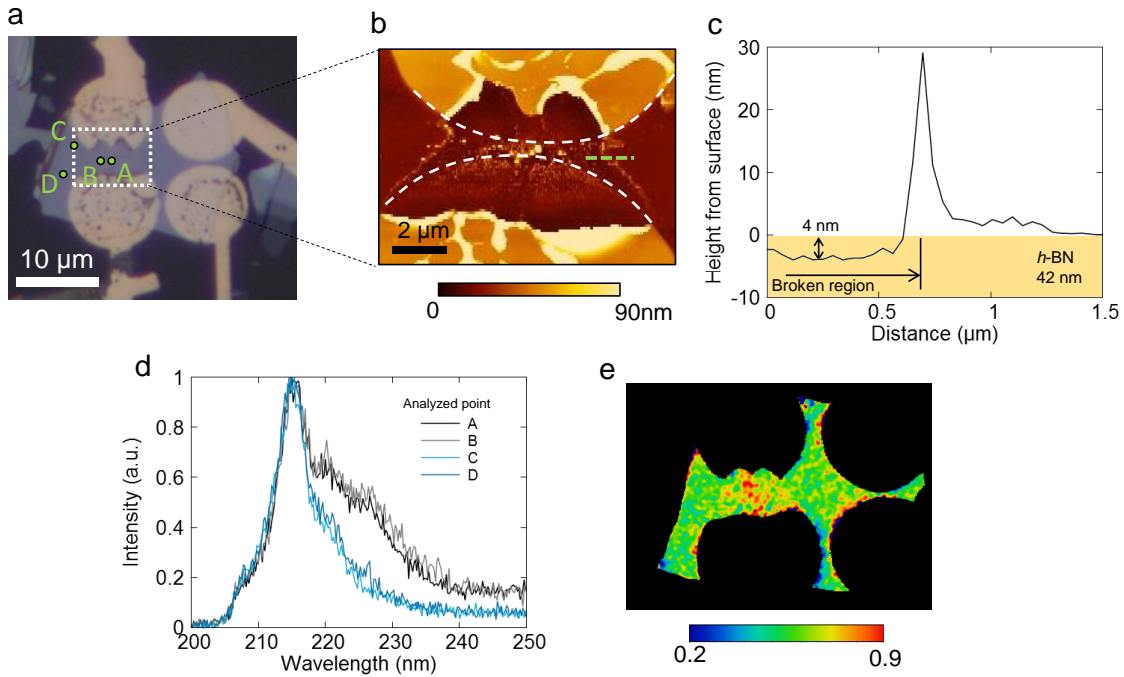

**Figure 3.** Characterization of the remaining *h*-BN layers after breakdown. (a) Optical microscopy image of *h*-BN after dielectric breakdown. The electrodes were broken because of the high current density at breakdown. (b) AFM image corresponding to the region indicated by the white-dashed rectangle in **a**. The white broken lines indicate the edges of the electrodes before breakdown. (c) Height profile along the green-dashed line in **b**. The heights of the *h*-BN in the channel and below the metal electrode were reduced by 4 nm. (d) CL spectra normalized to the intensity at 215 nm, which corresponds to free excitons in *h*-BN. The spectra were obtained at the four points labeled A-D in **a**. The broad peak near 225 nm that appears in the severely fractured regions A and B is attributed to stacking faults. (e) CL mapping for the intensity ratio between 220 nm and 215 nm, which was obtained through image processing using a 3 × 3 Gaussian kernel, which is used as a low-pass filter.

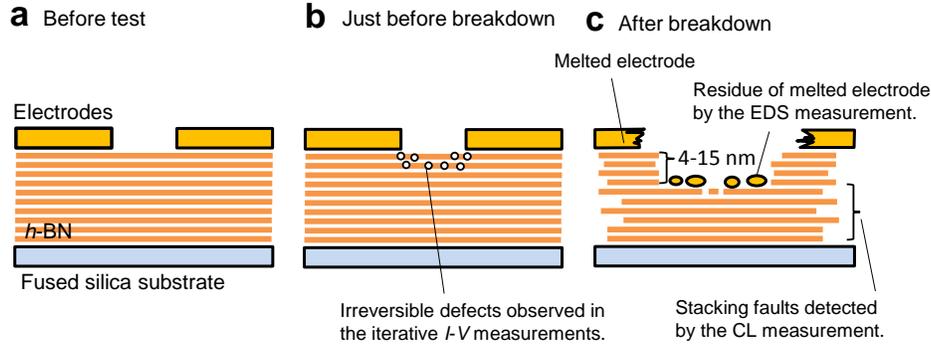

**Figure 4.** Schematic illustrations of the fracturing process. (**a**) *h*-BN before a breakdown test. (**b**) Defect formation just before breakdown. Open circles indicate irreversible defects. (**c**) Fracturing after breakdown. Several upper *h*-BN layers and the metal electrodes are thermally melted because of the high current density. Stacking faults are mechanically formed by the impact at breakdown.

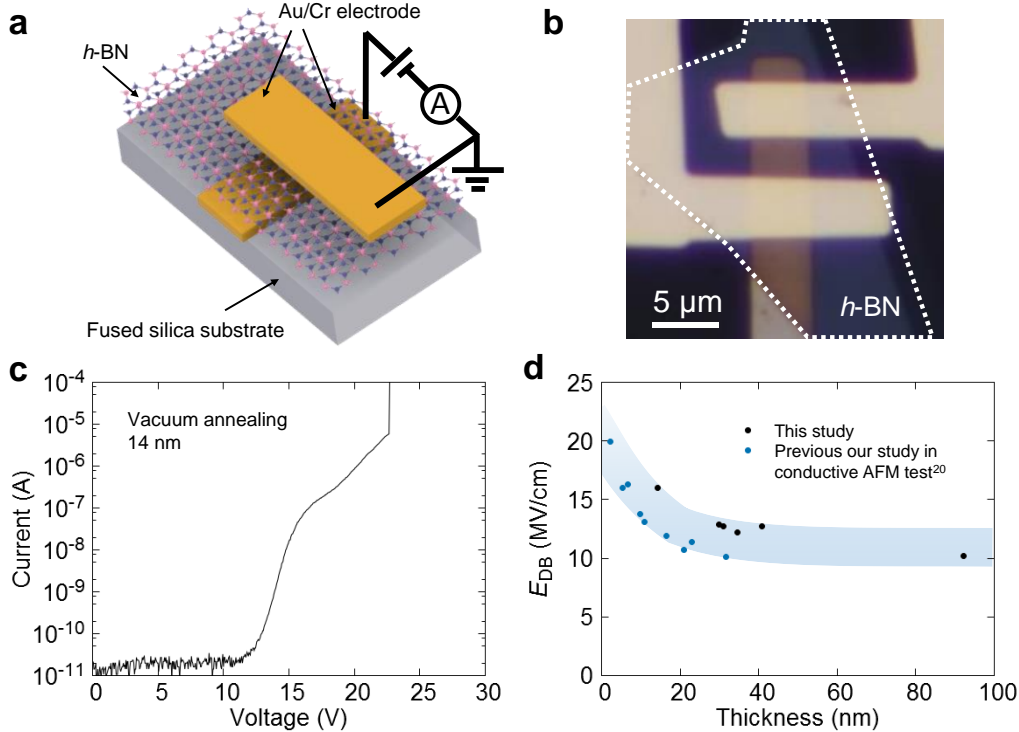

**Figure 5.** Dielectric breakdown in the direction parallel to the *c* axis. (**a**) Schematic drawing of a typical *h*-BN device. (**b**) Optical microscopy image of a fabricated *h*-BN device. (**c**) *I-V* curve for *h*-BN with a thickness of 14 nm. (**d**) $E_{BD//c}$ as a function of the *h*-BN thickness.

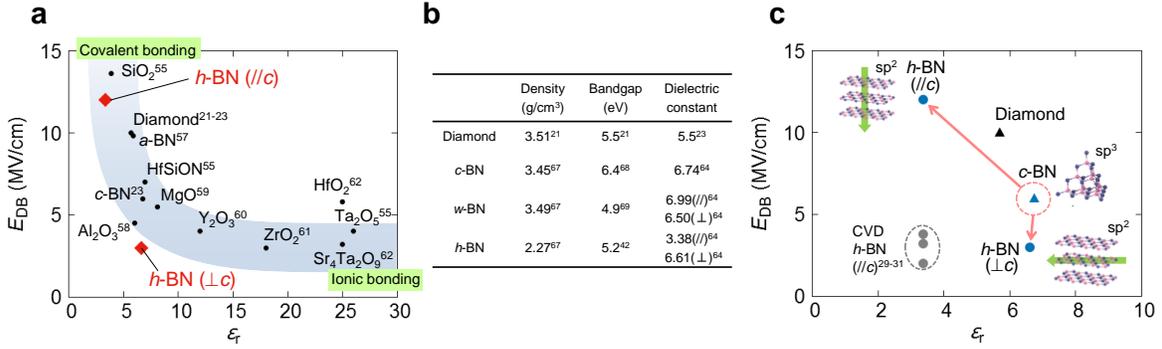

**Figure 6.** General trend of $E_{BD}$ vs $\varepsilon_r$ for various materials. (a) $E_{BD}$ as a function of $\varepsilon_r$ for various materials, along with the present $h$-BN data. Conventional insulators exhibit an empirical inverse relation between $E_{BD}$ and $\varepsilon_r$. (b) Physical properties of three polymorphs for BN. $h$-BN exhibits strong anisotropy in the dielectric constant. (c) $E_{BD}$ as a function of $\varepsilon_r$ for $h$-BN, $c$-BN and diamond.

## DISCUSSION

First, let us discuss the anisotropy of $E_{BD}$ in $h$-BN. Empirically, $E_{BD}$ has been found to be inversely related to the dielectric constant ($\varepsilon_r$) in well-studied insulators, as shown in **Fig. 6a**.[20-22,43-50] Insulators with strong covalent bonding have higher $E_{BD}$ and lower $\varepsilon_r$, whereas those with ionic bonding exhibit larger $\varepsilon_r$ because of their higher polarization. $h$-BN exhibits strong anisotropy in $\varepsilon_r$ with $\varepsilon_{r//c}$ and $\varepsilon_{r\perp c}$ of 3.38 and 6.61, respectively.[51-53] Because of this anisotropy in $\varepsilon_r$, $h$-BN follows the empirically established relationship, even though the difference between $E_{BD//c}$ and $E_{BD\perp c}$ is large. Here, the origin of the anisotropy is discussed based on the energy dispersion of the electrons in $h$-BN, because this empirical relationship can be simply converted into that of the band gap energy vs. $\varepsilon_r$.[54] In the case of conventional amorphous oxides, when an approximately random injection of electrons from the metal electrode into the conduction band of amorphous oxides is considered, $e$-$h$ pair formation in the amorphous oxides will be macroscopically determined by the fundamental band gap, which is the band gap between the conduction band minimum and the valence band maximum due to the scattering at the metal/$h$-BN interface and during the FN tunneling in the amorphous oxides. On the other hand, because the present $h$-BN sample is a high-quality single crystal, the diffuse scattering during the FN tunneling will be ignored. Based on the clear observation of a large anisotropy in $E_{BD}$, for simplicity, it may be possible to assume that the dielectric breakdown is mainly determined by the "first" $e$-$h$ pair formation in $h$-BN. When electrons are injected by the electrical field and tunnel into the conduction band of $h$-BN without scattering,[55,56] the $e$-$h$ pair formation could be strongly dependent on the band gap at different directions. Here, the band gap at the A point ($//c$) is 10.6 eV, whereas the band gap at the K or M points ($\perp c$) is ~5.9 eV.[57] Based on this discussion, the anisotropy of $E_{BD}$ in $h$-BN could be due to the large anisotropy in $\varepsilon_r$, or, in other words, a large anisotropy in the band gap.

Next, to understand the entire picture of anisotropy in $E_{BD}$, let us consider the crystal structure change from sp$^3$ of cubic-BN ($c$-BN) to sp$^2$ of $h$-BN, as shown in **Fig. 6c**. The physical properties for the two main polymorphs, $h$-BN and $c$-BN, are tabulated in **Fig. 6b**, together with that of $w$-BN.[20,22,31,51,52,58,59] $E_{BD\perp c}$ for $h$-BN becomes smaller than that for $c$-BN, while $E_{BD//c}$ for $h$-BN drastically increases. Therefore, the relatively high $E_{BD}$ concentrated only in the direction parallel to the $c$-axis is made possible by conceding a weak bonding direction in the highly anisotropic crystal structure. Therefore, it should be emphasized that the value of $E_{BD//c}$ = ~12 MV/cm for $h$-BN becomes larger than the ideal value of ~10 MV/cm for diamond. A similar situation may be expected for mica, which also has a quite high $E_{BD//c}$ (~10 MV/cm),[3,4] even though there are no available experimental data for $E_{BD\perp c}$.

Finally, let us discuss the relationship between $E_{BD}$ and $\varepsilon_r$ for $h$-BN from the perspective of device applications. The advantage of $h$-BN as a gate insulator is its high $E_{BD//c}$, whereas the disadvantage is its low $\varepsilon_{r//c}$. These characteristics constitute a trade-off relation.[60] To overcome this issue, combining $h$-BN with dielectrics with a high dielectric constant is the key to reducing the effective oxide thickness.[61,62] In terms of scalability, CVD or other techniques for the large-area $h$-BN is an important issue for electron device applications based on 2D materials. However, common problems such as the impurities, the presence of the wrinkle and grain boundaries and the nonuniformity of thickness[63] result in the lower $E_{BD}$ than that of the present sample exfoliated from the single crystal, as plotted in **Fig. 6c**.[41,64,65] The presented $E_{BD//c}$ value can be used as the standard value for confirming the quality of CVD-grown $h$-BN.[41,42,63-70]

## CONCLUSIONS

A thorough study conducted with careful attention to the relative humidity and the possible current paths in the device structure revealed an anisotropic $E_{BD}$ in single-crystal exfoliated $h$-BN. $E_{BD\perp c}$ was measured to be 3 MV/cm in the investigated device structure. According to analyses based on AFM, EDS, and CL measurements performed after dielectric breakdown, the expected fracturing process is as follows: the metal electrodes and several upper layers of $h$-BN are thermally melted because of the high current density that forms at breakdown after the irreversible defect formation, and the impact of the breakdown creates stacking faults in the remaining $h$-BN layers. By contrast, the value of $E_{BD//c}$ measured in an MIM structure was 12 MV/cm. When the crystal structure is changed from sp$^3$ of cubic-BN ($c$-BN) to sp$^2$ of $h$-BN, $E_{BD\perp c}$ for $h$-BN becomes smaller than that for $c$-BN, while $E_{BD//c}$ for $h$-BN drastically increases. Therefore, $h$-BN can possess a relatively high $E_{BD}$ concentrated only in the direction parallel to the $c$ axis by conceding a weak bonding direction in the highly anisotropic crystal structure. This explains the origin of the observation that the $E_{BD//c}$ for $h$-BN is higher than that for diamond. From the perspective of 2D device applications, it is suggested that $E_{BD//c}$ is sufficiently high and that the combination of $h$-BN with dielectrics of a high dielectric is the key to overcoming the disadvantage of the low $\varepsilon_{r//c}$. Moreover, the presented $E_{BD}$ value can be regarded as the standard for qualifying the crystallinity of $h$-BN layers grown via CVD.

## SUPPORTING INFORMATION
Details of the thickness measurement; Numerical calculation of the electrostatic field around the electrode tip; Photograph of the experimental set up; $I$-$V$ curve for the fused silica substrate; Sequential images of the light emitted at the breakdown; Iterative $I$-$V$ measurement; EDS analysis; Optical and AFM images for the fused silica substrate and $h$-BN after the breakdown. This material is available free of charge via the Internet at http://pubs.acs.org.

## AUTHOR INFORMATION

**Corresponding Author**
Email: *hattori@adam.t.u-tokyo.ac.jp,
**nagashio@material.t.u-tokyo.ac.jp

**Acknowledgement**
This research was partly supported by JSPS KAKENHI Grant Numbers JP25107004, JP16H04343, JP16K14446, & JP26886003.

**Notes**
The authors declare no competing financial interests.

# Supporting Information

# Anisotropic Dielectric Breakdown Strength of Single Crystal Hexagonal Boron Nitride

*Yoshiaki Hattori*[*†], *Takashi Taniguchi*[‡], *Kenji Watanabe*[‡] *and Kosuke Nagashio*[**†§]

[†]Department of Materials Engineering, The University of Tokyo, Tokyo 113-8656, Japan

[‡]National Institute of Materials Science, Ibaraki 305-0044, Japan

[§]PRESTO, Japan Science and Technology Agency (JST), Tokyo 113-8656, Japan

*hattori@adam.t.u-tokyo.ac.jp, **nagashio@material.t.u-tokyo.ac.jp

**Note 1: Concentration effect of the electrical field for $E_{BD\perp c}$.** In the following figure, the present device structure is compared with the ideal one in terms of the concentration effect of the electrical field. Because it is very difficult to fabricate the ideal device structure from the experimental viewpoint, the present device structure is quite general for a thin film. When the present structure is selected, the electrical field is concentrated in the edge region of electrode, as shown in **Fig. 1c** (dotted white circle). In general, the concentration effect of the electrical field decreases the breakdown voltage. Then, the shape factor is defined as α in the equation of $E_{BD}^{ideal}$ = α $E_{BD}^{present}$, where $E_{BD}$ is defined by $V_{BD}/L_{gap}$ here. We have not clarified the shape factor for the present structure yet. However, the shape factor has been historically studied. In these studies, the shape factor was less than ~1.5 even for the device in which the sample is sandwiched by pin electrode.[S1-S5] This pin electrode case provides the highest concentration effect. However, because of old papers, the scale of the sample size is considerably large, where the thick sample over 1 μm was broken by high voltage over 1000 V. It may not be reasonable to directly apply the old references to the present situation. Here, for our $h$-BN breakdown test for $E_{BD//c}$ in **Fig. 5d**, the difference in $E_{BD//c}$ between "plate-plate electrode" and "pin-plate electrode" is negligible, where pin-plate case was obtained by conductive-AFM with a radius of curvature of ~100 nm. From above discussion, the concentration effect of the electrical field for $E_{BD\perp c}$ is expected to be small enough in this study. Therefore, the raw data for $E_{BD\perp c}$ was plotted in **Fig. 6a**. Note that even if we apply α =1.5 for $h$-BN, the relation of $E_{BD\perp c} >> E_{BD//c}$ is still valid.

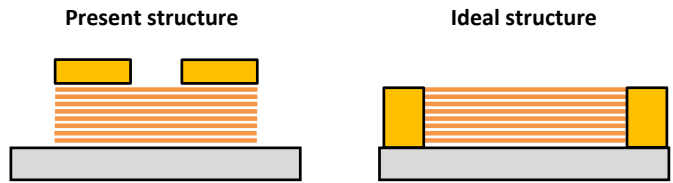

**References:**

S1. Mason, J. H. Breakdown of Solid Dielectrics in Divergent Fields. *Proc. IEEE-Part C: Monographs* **1955**, *102*, 254–263.

S2. Kuffel, E. Influence of Humidity on the Breakdown Voltage of Sphere-Gaps and Uniform-Field Gaps. *Proc. IEEE-Part A: Power Eng.* **1961**, *108*, 295–301.

S3. Akahira, T.; Gemant, A. Elektrische Festigkeit Mechanisch Beanspruchter Isolierstoffe. *Arch. Elektrotech.* **1933**, *27*, 577–585.

S4. Schwaiger, A. Über die Ermittlung der Durchschlagfestigkeit von hygroskopischen Isoliermaterialien. *Arch. Elektrotech.* **1915**, *3*, 332–344.

S5 Sorge, J. Über die elektrische Festigkeit einiger flüssiger Dielektrika. *Arch. Elektrotech.* **1924**, *13*, 189–212.

**Note 2: Transfer method.** The devices for the $E_{BD//c}$ measurements were fabricated using a transfer technique. A poly(methyl methacrylate) (PMMA) layer with a thickness of 1 μm was spin-coated on a 200-μm-thick polypropylene sheet. $h$-BN flakes were mechanically exfoliated onto it. Metal strips (15-nm Cr / 15-nm Au) with a width of 5 μm were fabricated on the fused silica via EB lithography. Then, the $h$-BN flakes on the PMMA/polypropylene sheet were transferred to the metal strips on the substrate. After the removal of the PMMA with acetone and isopropyl alcohol, the top electrodes (15-nm Cr / 30-nm Au) were patterned on the $h$-BN via EB lithography.

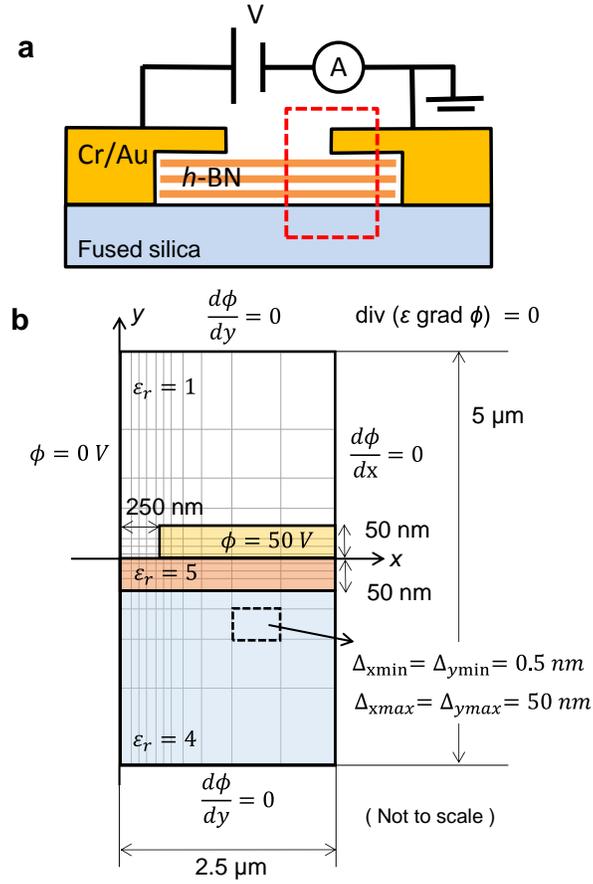

**Figure S1:** (**a**) Schematic diagram of the measurement setup. (**b**) 2D simulation model.

**Numerical calculation of the electrical field.** The electrostatic field around the tip of electrode is calculated. The potential ($\phi$) around the electrodes satisfies the following Poisson's equation,

$$\text{div}\,(\varepsilon\,\text{grad}\,\phi) = 0,$$

where $\varepsilon$ is the dielectric constant. Relative dielectric constant $\varepsilon_r$ = 5 and 4 are used for *h*-BN and fused silica, respectively. The electrostatic field (*E*) is calculated as – grad $\phi$. The governing equation can be solved explicitly using the finite-volume method in a simplified 2D model (**Figure S1a, b**) with a Cartesian geometry (*x*, *y*). The calculation area is divided into 129×79 non-uniform grids. The minimum and maximum grid size is 0.5 nm and 50 nm, respectively. As a matter of calculation convenience, only one side of area is calculated. Thereby left surface is grounded and the electrode potential is set up to half of the actually applied voltage. The boundary conditions except for the left surface are Neumann boundary.

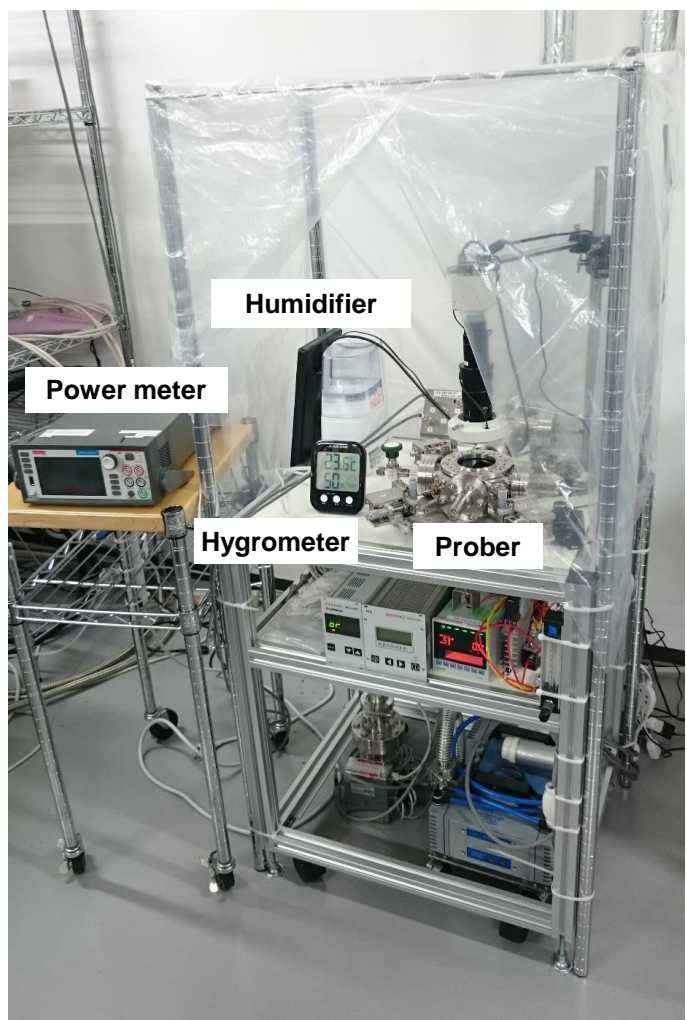

**Figure S2:** Experimental set up for the *I-V* measurements under relative humidity control. In order to keep the relative humidity stable during the measurement, the measurement area was separated with the transparent plastic sheet, in which the humidifier was placed. It should be noted that the relative humidity in the whole laboratory room was also controlled by combing the humidifier and the dehumidifier, because the local humidity above the sample is quite sensitive to the convection.

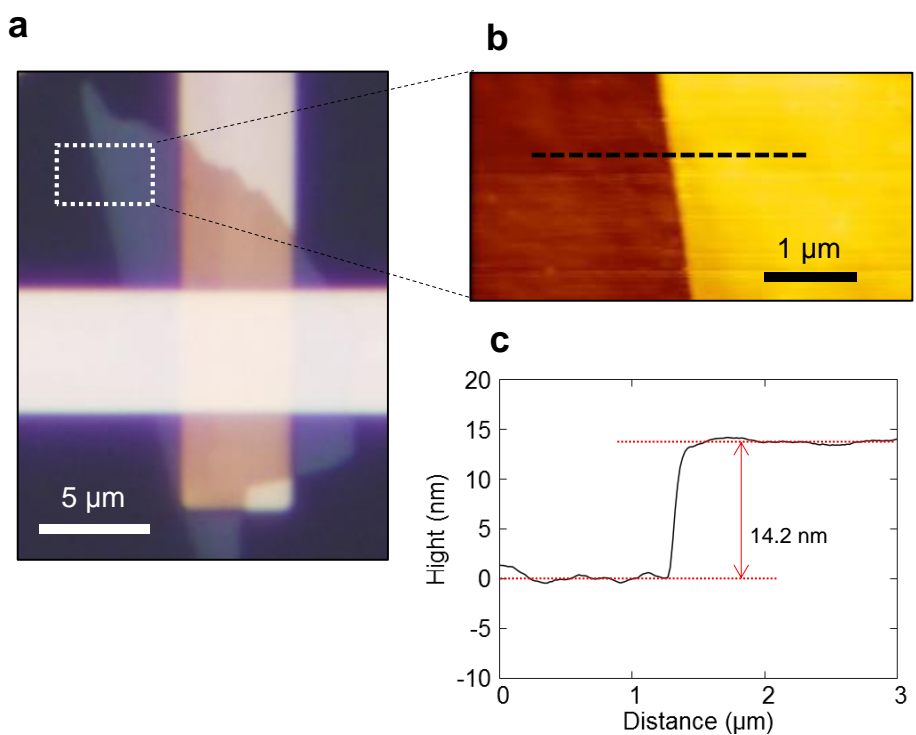

**Figure S3:** Thickness measurement of *h*-BN. (**a**) Optical image of the fabricated device. (**b**) AFM image of the white-dashed rectangular region in **a**. (**c**) Height profile along the dashed line in **b**.

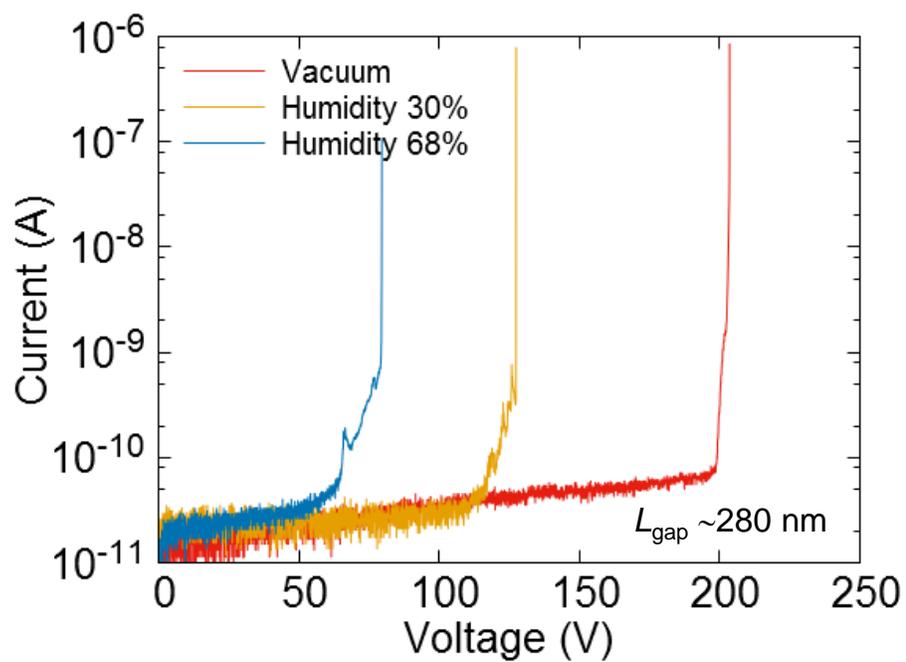

**Figure S4:** *I-V* curve for the fused silica substrate for the $L_{gap}$ ~280 nm.

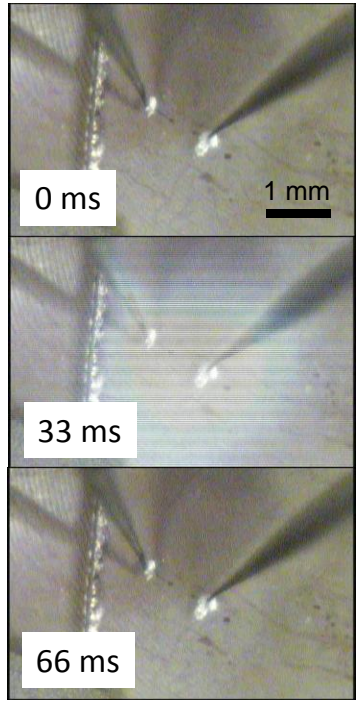

**Figure S5:** Sequential images of the light emitted at the dielectric breakdown. These images were taken by conventional CCD camera (STC-630CT, Sentech) with a frame rate of 33 ms. The light was observed also by naked eyes.

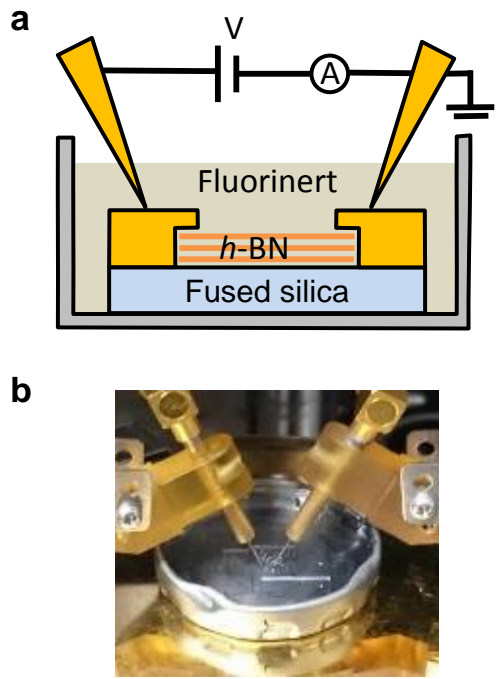

**Figure S6:** Schematic drawing (**a**) and photograph (**b**) of the experimental setup in the insulator oil (3M Fluorinert FC-3283).

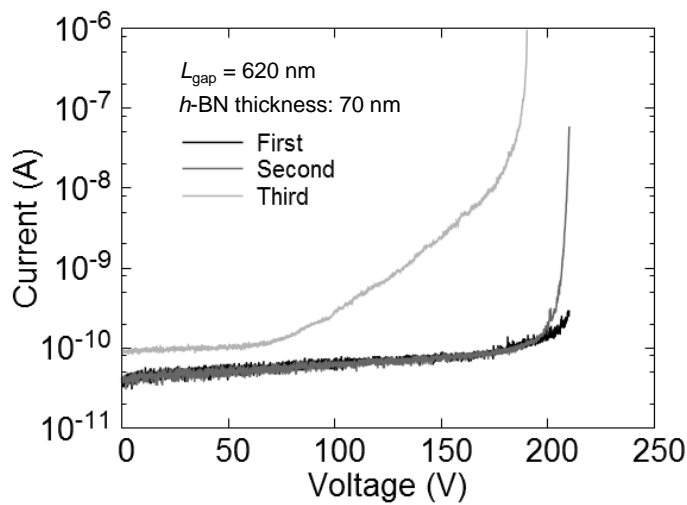

**Figure S7:** Iterative *I-V* measurement for *h*-BN with a thickness of 70 nm. The device with $L_{gap}$ = 620 nm was measured three times successively. Applied voltages is the range from 0 V to 210 V, which is the maximum of the source meter. In the first measurement the current gradually increases from 180 V. However, the breakdown which causes the catastrophic fracture for both *h*-BN and electrode didn't occur. The iterative measurement was conducted to observe the degradation of *h*-BN before the breakdown voltage. In the second measurement, although the increase in leakage current was observed over 200 V, no phenomena related with the breakdown such as the perceivable light emission, the melted electrode, and the removal of *h*-BN were observed at all. In the third measurement, the leakage current was detected at the early stage of the applied voltage and further increased drastically over 70 V. These observation suggests that the irreversible defects were formed before the breakdown.

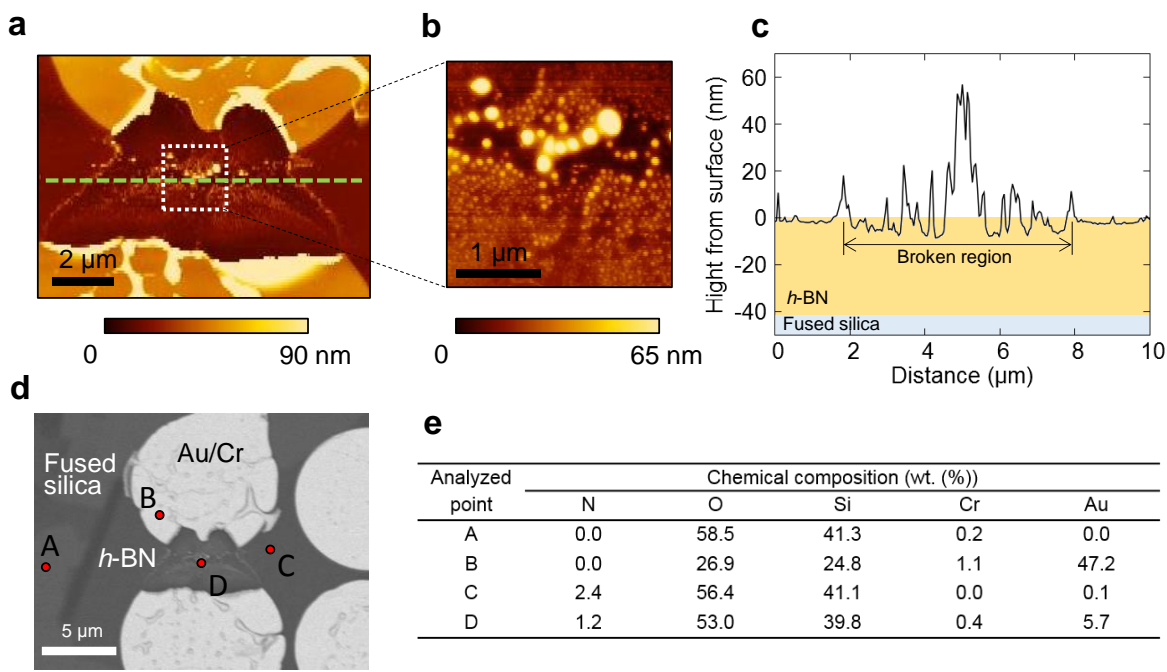

**Figure S8:** AFM image and energy dispersive X-ray spectrometry (EDS) analysis for the remaining layers after the breakdown. (**a**) AFM image same as **Fig. 3b**. (**b**) Magnified image of the white-dashed rectangular region in **a**. (**c**) Height profile along the green-dashed line in **a**. (**d**) SEM image showing the positions A–D analyzed by EDS. (**e**) Chemical composition for positions A–D. The positions A and B are the fused silica substrate and Au/Cr electrode, respectively. The position C is unbroken region of $h$-BN, where no Au is detected. The position D is fractured region of h-BN, where Au is clearly detected. This suggests that the small spheres observed in **b** is Au electrode melted at the breakdown.

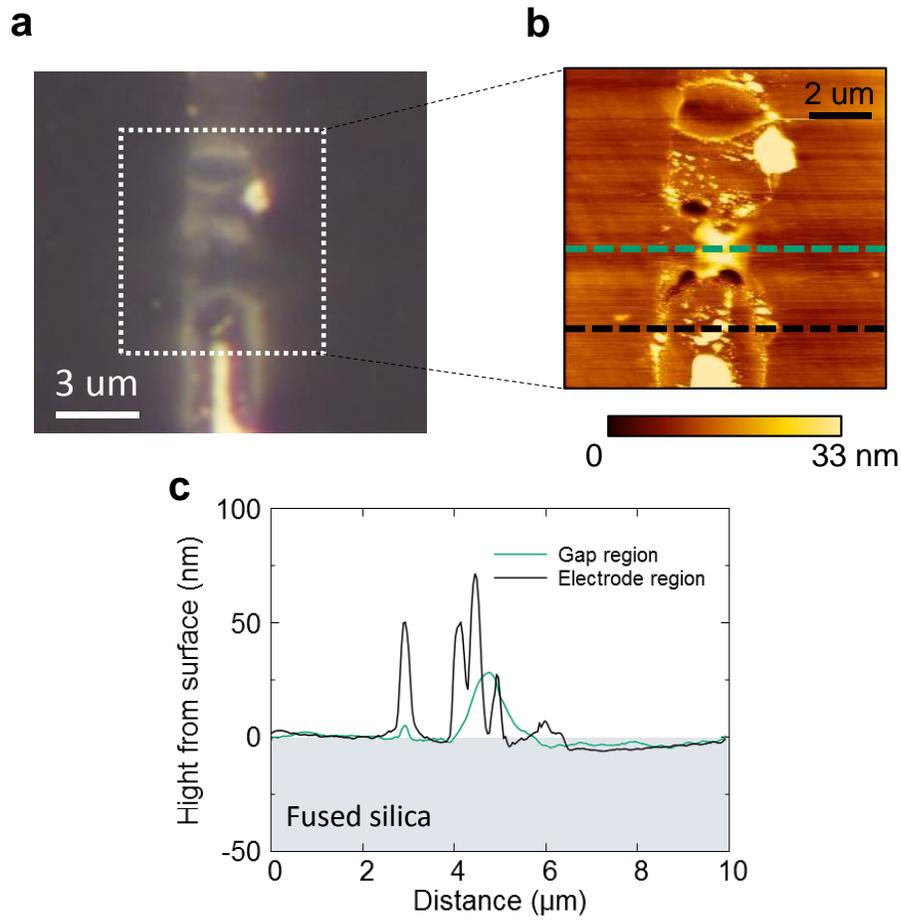

**Figure S9:** (**a**) Optical image of the fused silica substrate surface after the breakdown. The metal electrode was fabricated directly on a fused silica substrate without $h$-BN to evaluate the $E_{BD}$ of the fused silica substrate. The two electrodes disappeared due to the impact of the breakdown. (**b**) AFM image of the white-dashed rectangular region in **a**. (**c**) Height profile of the green and black dashed line in **b**. The gap region (green) was protruded locally. On the other hand, the fused silica substrate below the metal electrode has not been fractured at the breakdown, unlike $h$-BN, even though many residue of electrodes can be seen.

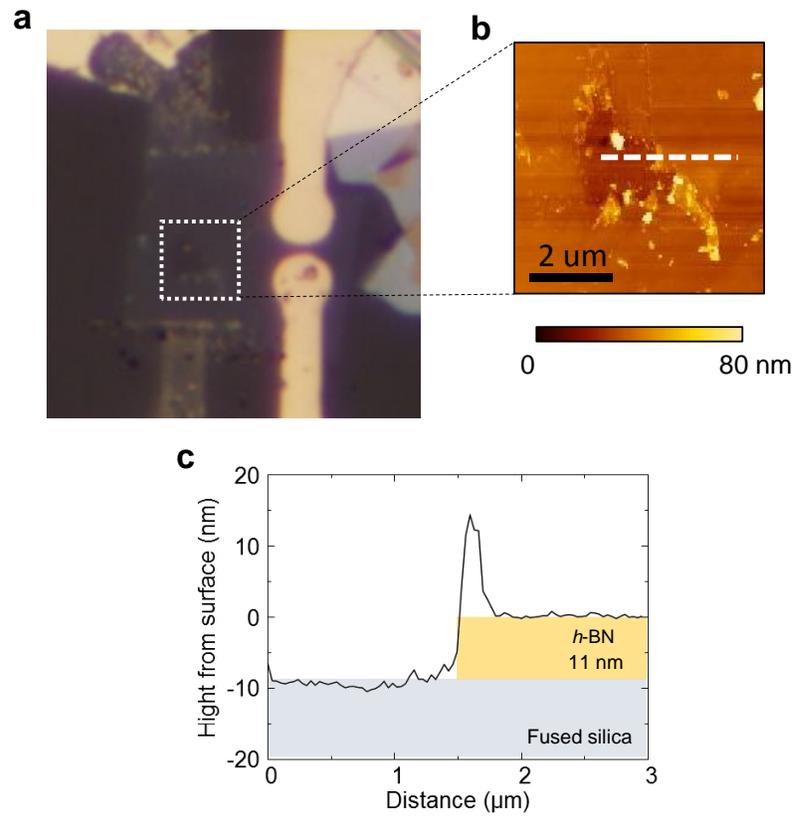

**Figure S10:** (**a**) Optical image of the surface of the thin *h*-BN (11-nm) with $L_{gap}$ = 440 nm after the breakdown. (**b**) AFM image of the white-dashed rectangular region in **a**. (**c**) Height profile of the white-dashed line in **b**. All the *h*-BN layers were melted due to the high current density at the breakdown. The damage in the fused silica substrate is not observed, suggesting that the current path of the *h*-BN/substrate interface is not the dominant in the breakdown.